\documentclass[a4paper]{jpconf}
\usepackage{amsmath}
\usepackage{epsfig}
\usepackage{amssymb}
\usepackage{graphicx}
\usepackage{color}
\definecolor{rosso}{cmyk}{0,1,1,0.4}
\definecolor{rossos}{cmyk}{0,1,1,0.55}
\definecolor{rossoc}{cmyk}{0,1,1,0.2}
\definecolor{blu}{cmyk}{1,1,0,0.3}
\definecolor{blus}{cmyk}{1,1,0,0.6}
\definecolor{bluc}{cmyk}{1,1,0,0.1}
\definecolor{verde}{cmyk}{0.92,0,0.59,0.25}
\definecolor{verdec}{cmyk}{0.92,0,0.59,0.15}
\definecolor{verdes}{cmyk}{0.92,0,0.59,0.4}

\def\bi#1{#1}
\def\circa#1{\,\raise.3ex\hbox{$#1$\kern-.75em\lower1ex\hbox{$\sim$}}\,}

\def\half{\frac{1}{2}} 
\def\ra{\rightarrow}

\newcommand{\beq}{\begin{equation}}
\newcommand{\eeq}{\end{equation}}
\newcommand{\bea}{\begin{eqnarray}}
\newcommand{\eea}{\end{eqnarray}}
\newcommand{\ba}{\begin{array}}
\newcommand{\ea}{\end{array}}
\newcommand{\bn}{\begin{enumerate}}
\newcommand{\en}{\end{enumerate}}
\newcommand{\bc}{\begin{center}}
\newcommand{\ec}{\end{center}}

\newcommand{\gsim}{\lower.7ex\hbox{$\;\stackrel{\textstyle>}{\sim}\;$}}
\newcommand{\baz}{\begin{array}{cc}}
\newcommand{\bad}{\begin{array}{ccc}}
\def\gtap{\mathrel{ \rlap{\raise 0.511ex \hbox{$>$}}{\lower 0.511ex
   \hbox{$\sim$}}}} 
\def\ltap{\mathrel{ \rlap{\raise 0.511ex
   \hbox{$<$}}{\lower 0.511ex \hbox{$\sim$}}}}
   \newcommand{\deltaatm}{\mbox{$\Delta m^2_{\mathrm{A}}$}}
   \newcommand{\deltasol}{\mbox{$ \Delta m^2_{\odot}$}}
\newcommand{\dmsol}{\mbox{$\Delta m^2_{\odot}$}}
\newcommand{\dma}{\mbox{$\Delta m^2_{\rm A}$}}

\newcommand{\pmns}{\mbox{$ U$}}

\begin{document}

\title{CP Violation and Lightest Neutrino Mass Effects in Thermal Leptogenesis}
\author{E Molinaro $^{a,}$\footnote{Talk given by E. Molinaro at DISCRETE'08, 11-16 December 2008, IFIC, Valencia, Spain.}, S T Petcov $^{a,b}$,T Shindou $^c$, Y Takanishi $^d$}
\address{$^a$ SISSA and INFN-Sezione di Trieste, Trieste I-34014, Italy}
\address{$^b$ Institute of Nuclear Research and Nuclear Energy, Bulgarian Academy 
of Sciences, 1784 Sofia, Bulgaria.}
\address{$^c$ DESY, Theory Group, Notkestrasse 85, D-22603 Hamburg, Germany}
\address{$^d$ Technische Universit\"at M\"unchen, Physik Department T31, James-Franck-Str. 1, D-85747 Garching, Germany}

\begin{abstract}
Effects of the lightest neutrino mass in ``flavoured''
leptogenesis when the CP-violation
necessary for the generation of the baryon asymmetry of the Universe
is due exclusively to the Dirac and/or Majorana phases in the neutrino
mixing matrix $U$ are discussed. The type I see-saw scenario with three heavy
right-handed Majorana neutrinos having hierarchical spectrum is
considered.  The ``orthogonal'' parametrisation of the matrix of
neutrino Yukawa couplings, which involves a complex orthogonal matrix
$R$, is employed. Results for light neutrino mass spectrum with normal
and inverted ordering (hierarchy) are reviewed. 
\end{abstract}

\section{Introduction}

We discuss the effects of the lightest neutrino mass in thermal leptogenesis \cite{FY,kuzmin}
where lepton flavor dynamics \cite{Barbieri99}$-$\cite{antusch} plays an important role in the generation of the observed baryon asymmetry of the Universe and the CP-violation required for the baryogenesis mechanism to work is due exclusively to the Dirac and/or Majorana CP-violating phases in the Pontecorvo-Maki-Nakagawa-Sakata (PMNS)~\cite{BPont57} neutrino mixing matrix. A detailed investigation of these effects was performed in reference \cite{MPST}. We review some of the results obtained in \cite{MPST}.  

The minimal scheme in which leptogenesis can be implemented is the
non-supersymmetric version of the type I see-saw~\cite{seesaw} model with two or
three heavy right-handed (RH) Majorana neutrinos. Taking into account the lepton flavour effects in
leptogenesis it was shown \cite{PPRio106} (see also \cite{Nielsen02,BrancoJ06,SBDibari06}) that if the
heavy Majorana neutrinos have a hierarchical spectrum, the observed baryon
asymmetry $Y_B$ can be produced even if the only source of
CP-violation is the Majorana and/or Dirac phase(s) in the PMNS
matrix $U_{\rm PMNS}\equiv U$. In this case the predicted value of the
baryon asymmetry depends explicitly (i.e.  directly) on $U$ and on the
CP-violating phases in $U$.  The results quoted above were
demonstrated to hold both for normal hierarchical (NH) and inverted
hierarchical (IH) spectrum of masses of the light Majorana neutrinos.  
In both these cases they were obtained
for negligible lightest neutrino mass and CP-conserving elements of
the orthogonal matrix $R$, present in the ``orthogonal''
parametrisation \cite{Casas01} of the matrix of neutrino Yukawa
couplings. The CP-invariance constraints imply that
the matrix $R$ could conserve the CP-symmetry if its elements are real
or purely imaginary \footnote{The more general case in which CP-violation arises from the combined effect between the ``low energy'' Majorana and/or Dirac phases in $U_{\rm PMNS}$ and the ``high energy'' CP-violating phases in a complex orthogonal matrix $R$, in thermal ``flavoured'' leptogenesis scenario, was addressed in \cite{Molinaro:2008rg} and \cite{Molinaro:2008cw}.}. We remark that for a CP-conserving matrix
$R$ and at temperatures $T\sim M_1\gtap 10^{12}$ GeV, the lepton flavours are indistinguishable
(one flavour approximation) and the total CP-asymmetry is always zero. 
In this case no baryon asymmetry is produced. 
One can prove \cite{PPRio106} that, for NH
spectrum and negligible lightest neutrino mass $m_1$ 
successful thermal leptogenesis can be realised for a real matrix $R$.  In contrast, in the
case of IH spectrum and negligible lightest neutrino mass ($m_3$), the
requisite baryon asymmetry was found to be produced for CP-conserving
matrix $R$ only if certain elements of $R$ are purely imaginary: for
real $R$ the baryon asymmetry $Y_B$ is strongly suppressed 
and leptogenesis cannot be successful for $M_1 \ltap
10^{12}~{\rm GeV}$ (i.e. in the regime in which the lepton flavour
effects are significant \cite{davidsonetal,niretal,davidsonetal2}).  

 In the present article we discuss the effects of the lightest
neutrino mass on ``flavoured'' (thermal) leptogenesis. We consider
the case when the CP-violation necessary for the generation of the
observed baryon asymmetry of the Universe is due exclusively to the
Dirac and/or Majorana CP-violating phases in the PMNS matrix $U$.  The
results we review correspond to the simplest type I see-saw scenario with
three heavy RH Majorana neutrinos $N_j$, $j=1,2,3$. The latter are
assumed to have a hierarchical mass spectrum, $M_1 \ll M_{2,3}$.
As a consequence, the generated 
baryon asymmetry $Y_B$ depends linearly on the mass of $N_1$, $M_1$, and
on the elements $R_{1j}$ of the matrix $R$, $j=1,2,3$, present in the 
orthogonal parametrisation of neutrino Yukawa couplings of $N_1$.
As was already mentioned previously, this
parametrisation involves an orthogonal matrix $R$, $R^TR = RR^T = {\bf
  1}$.  Although, in general, the matrix $R$ can be complex, i.e.
CP-violating, in \cite{MPST} we were primarily interested in the
possibility that $R$ conserves the CP-symmetry. We report results of the two
types of light neutrino mass spectrum allowed by the data \cite{STPNu04}: 
i) with normal ordering ($\deltaatm >0$), $m_1 < m_2
< m_3$, and ii) with inverted ordering ($\deltaatm < 0$), $m_3 < m_1 <
m_2$.  The case of inverted hierarchical (IH) spectrum and real (and
CP-conserving) matrix $R$ is reviewed in detail
\footnote{New material, not published in \cite{MPST}, is presented in figures 2 and 3.}. 
 
The analysis in \cite{MPST} was performed for negligible renormalisation group (RG)
running of $m_j$ and of the parameters in the PMNS matrix $\pmns$ from
$M_Z$ to $M_1$.  This possibility is realised for sufficiently small values of the lightest
neutrino mass ${\rm min}(m_j)$~\cite{rad,PST06}, e.g., for ${\rm
  min}(m_j) \ltap 0.10$ eV.  The latter condition is fulfilled for the
NH and IH neutrino mass spectra, as well as for spectrum with partial
hierarchy \cite{BPP1}. Under the indicated condition
$m_j$, and correspondingly $\deltaatm$ and $\deltasol$, and $U$ can be
taken at the scale $\sim M_Z$, at which the neutrino mixing parameters
are measured.

\section{``Low Energy'' CP-Violation and CP-Asymmetry in Flavoured Leptogenesis}

Following the discussion in \cite{PPRio106}, we introduce  combinations
between the elements of the neutrino mixing matrix $U$ and 
the orthogonal matrix $R$ that appears in the Casas-Ibarra parametrisation \cite{Casas01}
of the matrix of neutrino Yukawa couplings:
\begin{equation}\label{P}
	P_{jkml}\equiv R_{j k}\,R_{j m} \, U^{\ast}_{lk} \, U_{lm}\,,~ k\neq m\,. 
\end{equation}
If CP-invariance doesn't hold, then one can easily prove that $P_{jkml}$ is not a real quantity:
\begin{equation}\label{CP-cons}
	{\rm Im}(P_{jkml}) \neq 0\,. 
\end{equation}
In the parametrisation of the neutrino Yukawa couplings considered, 
the condition (\ref{CP-cons}) triggers CP-violation in the thermal leptogenesis
scenario, when the dynamics of the flavour states plays a role in the generation 
of the baryon asymmetry of the Universe. In particular, from (\ref{P}) it is 
possible to understand what is the interplay between the ``low energy'' CP
violation, encoded in the Majorana and/or Dirac phases present in the 
neutrino mixing matrix $U$, and the ``high energy'' CP-violating phases of the
matrix $R$ and to disentangle the two contributions. In the present article we are primarily
interested in the situation in which the CP-violation necessary for having successful leptogenesis
can arise exclusively from ``low energy'' physics in the lepton sector.
For this reason, we impose the orthogonal matrix $R$ to satisfy CP-invariance constraints, i.e.  
all the matrix elements, $R_{ij}$, are real or purely imaginary. If this is the case,
CP-violation, and therefore condition (\ref{CP-cons}), is accomplished through the
the neutrino mixing matrix $U$ (``low energy'' CP-violation).
For real or purely imaginary $R_{1j}R_{1k}$, $j \neq k$,
the CP-asymmetries $\epsilon_{l}$ is given by{\footnotesize
\begin{eqnarray}
	\epsilon_{l}
	= -\, \frac{3M_1}{16\pi v^2} 
	\frac{\sum_k \sum_{j>k}\sqrt{m_k m_j}\,
	(m_j - m_k)\, \rho_{kj}|R_{1k}R_{1j}|\,
	{\rm Im}\,\left(U^*_{l k}\, U_{l j}\right)}
	{\sum_i m_i\, |R_{1 i}|^2},~{\rm Im}\,(R_{1k} R_{1j})=0
	\\
	\epsilon_{l} = -\, \frac{3M_1}{16\pi v^2} 
	\frac{\sum_k \sum_{j>k}\sqrt{m_k m_j}\,
	(m_j + m_k)\, \rho_{kj}|R_{1k} R_{1j}|\,
	{\rm Re}\,\left(U^*_{l k} U_{l j}\right)}
	{\sum_i m_i\, |R_{1 i}|^2},~{\rm Re}\,(R_{1k}R_{1j})=0
	\label{epsa3}
\end{eqnarray}}
\noindent
with $R_{1j}R_{1k} = \rho_{jk}~|R_{1j} R_{1k}|$ and
$R_{1j}R_{1k} = i\rho_{jk}~|R_{1j} R_{1k}|$, $\rho_{jk} = \pm 1$, $j\neq k$. 
Note that, according to condition (\ref{CP-cons}), real (purely imaginary) $R_{1k}R_{1j}$ and purely
imaginary (real) $U^*_{lk} U_{lj}$, $j\neq k$, implies violation of
CP-invariance by the matrix $R$ \cite{PPRio106}.
An interesting possibility is for example when the Dirac phase $\delta$
and the effective Majorana phases $\alpha_{31}$, $\alpha_{21} $\footnote{We use the standard parametrisation of the PMNS matrix as defined in \cite{PPRio106}.} take the CP-conserving values: $\delta=0$, $\alpha_{31}=\pi$
and $\alpha_{21}=0$. Then, for real $R_{1j}$, $j=1,2,3$, condition (\ref{CP-cons}) is fulfilled and 
the CP-asymmetry $\epsilon_l$ is different from zero. We say in this case that both the PMNS matrix and 
the orthogonal matrix are CP-conserving, but the neutrino Yukawa couplings still violate CP-symmetry.
More specifically, it is impossible to construct a high energy observable that is sensitive to
CP-symmetry breaking and depends only on the matrix $R$. The only possibility to 
break the symmetry at high energy scales is to couple the matrix $R$ to the PMNS
neutrino mixing matrix, as in $\epsilon_l$.

In order for the CP-symmetry to be broken at low energies, we should have both ${\rm Re}(U^*_{lk} U_{lj})\neq 0$ and ${\rm Im}(U^*_{lk} U_{lj})\neq 0$ (see \cite{PPRio106} for further details on this point).

\section{Light Neutrino Mass Spectrum with Inverted Ordering and Real $R_{1j}$}

The case of inverted hierarchical (IH) neutrino mass spectrum, $m_3
\ll m_1 < m_2$, $m_{1,2} \cong \sqrt{|\deltaatm|}$, is of particular
interest since, as was already mentioned, for real
$R_{1j}$, $j=1,2,3$, IH spectrum and negligible lightest neutrino mass
$m_3 \cong 0$, it is impossible to generate the observed baryon
asymmetry $Y_B \cong 8.6\times 10^{-11}$ in the regime of
``flavoured'' leptogenesis \cite{PPRio106}, i.e. for $M_1 \ltap
10^{12}~{\rm GeV}$, if the only source of CP-violation are the
Majorana and/or Dirac phases in the PMNS matrix. It can be proven that 
for $m_3=0$ and $R_{13}=0$, the resulting baryon asymmetry is always suppressed
by the factor $\deltasol/(2\deltaatm) \cong 1.6 \times 10^{-2}$.
We analyse the generation of the baryon asymmetry
$Y_B$ for real $R_{1j}$, $j=1,2,3$, when $m_3$ is non-negligible. We
assume that $Y_B$ is produced in the two-flavour regime,
$10^9~{\rm GeV}\ltap M_1 \ltap 10^{12}$ GeV.
Under these conditions
the terms $\propto \sqrt{m_3}$ in $\epsilon_{l}$ will be dominant
provided \cite{PPRio106}
\begin{equation}
2\left( \frac{m_3}{\sqrt{\dmsol}} \right)^{\half}
\left(\frac{\dma}{\dmsol} \right)^{\frac{3}{4}}
\frac{\left|R_{13}\right|}{\left|R_{11(12)}\right|}\gg 1\,. 
\label{limit}
\end{equation}
This condition can be easily satisfied if $R_{11}\ra 0$, or $R_{12}\ra 0$,
and if $m_3$ is sufficiently large. The neutrino mass spectrum 
is still hierarchical for $m_3$ having a value $m_3 \ltap 5\times
10^{-3}~{\rm eV} \ll \sqrt{|\deltaatm|}$. The general analysis is performed 
for values of $m_3$ from the interval $10^{-10}~{\rm eV}
\ltap m_3 \ltap 5\times 10^{-2}~{\rm eV}$.

\begin{figure}[t!!]
\begin{center}
\vspace{-1cm}
\includegraphics[width=13.5cm,height=9.5cm]{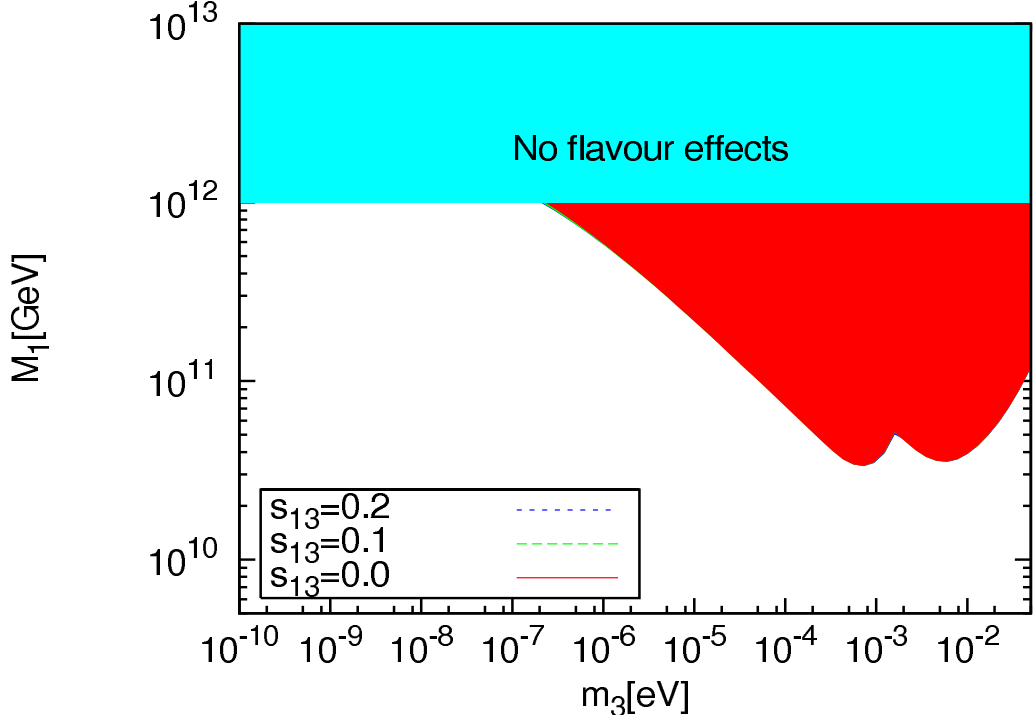}
\caption{
  \label{fig1IOM1m3} Values of $m_3$ and $M_1$ for which the
  ``flavoured'' leptogenesis is successful, generating baryon
  asymmetry $|Y_B| = 8.6\times 10^{-11}$ (red/dark shaded area). The
  figure corresponds to hierarchical heavy Majorana neutrinos, light
  neutrino mass spectrum with inverted ordering (hierarchy), $m_3 <
  m_1 < m_2$, and real elements $R_{1j}$ of the matrix $R$. The results shown are obtained
  using the best fit values of neutrino oscillation parameters:
  $\deltasol = 8.0\times 10^{-5}~{\rm eV}^2$, $\deltaatm = 2.5\times
  10^{-3}~{\rm eV}^2$, $\sin^2\theta_{12}=0.30$ and
  $\sin^22\theta_{23}=1$.}
\end{center}
\end{figure}

In Fig. 1 we show the correlated values of $M_1$ and $m_3$ for which
one can have successful leptogenesis in the case of neutrino mass
spectrum with inverted ordering and CP-violation due to the Majorana
and Dirac phases in $U_{\rm PMNS}$.  The figure was obtained by
performing, for given $m_3$ from the interval $10^{-10} {\rm eV}\leq m_3 \leq
0.05$ eV, a thorough scan of the relevant parameter space searching
for possible enhancement or suppression of the baryon asymmetry with
respect to that found for $m_3 = 0$.  The real elements of the
$R-$matrix of interest, $R_{1j}$, $j=1,2,3$, were allowed to vary in
their full ranges determined by the condition of orthogonality of the
matrix $R$: $R^2_{11} + R^2_{12} + R^2_{13} = 1$. The Majorana phases $\alpha_{21,31}$
were varied in the interval $[0,2\pi]$.  The calculations were
performed for three values of the CHOOZ angle $\theta_{13}$,
corresponding to $\sin\theta_{13} = 0;~0.1;~0.2$.  In the cases of
$\sin\theta_{13}\neq 0$, the Dirac phase $\delta$ was allowed to take
values in the interval $[0,2\pi]$. The heavy Majorana neutrino mass
$M_1$ was varied in the interval $10^{9}~{\rm GeV} \leq M_1 \leq
10^{12}$ GeV.  For given $m_3$, the minimal value of the mass $M_1$,
for which the leptogenesis is successful, generating $|Y_B| \cong
8.6\times 10^{-11}$, was obtained for the values of the other
parameters which maximise $|Y_B|$. We have found that in the case of IH spectrum with non-negligible
$m_3$, $m_3 \ll \sqrt{|\deltaatm|}$, the generated baryon asymmetry
$|Y_B|$ can be strongly enhanced in comparison with the asymmetry
$|Y_B|$ produced if $m_3 \cong 0$. The enhancement can be by a factor
of $\sim 100$, or even by a larger factor.  As a consequence, one can
have successful leptogenesis for IH spectrum with $m_3 \gtap 5\times
10^{-6}$ eV even if the elements $R_{1j}$ of $R$ are real and the
requisite CP-violation is provided by the Majorana or Dirac phase(s)
in the PMNS matrix. As a consequence, successful thermal leptogenesis is realised
for $5\times 10^{-6}~{\rm eV} \ltap m_3 \ltap
5\times 10^{-2}$ eV. The results of our analysis show that for Majorana
CP-violation from $U_{\rm PMNS}$, successful leptogenesis can be obtained
for $M_1\gtap 3.0\times 10^{10}~{\rm  GeV}$. Larger values of $M_1$ are typically required if
the CP-violation is due to the Dirac phase $\delta$: $M_1\gtap
10^{11}~{\rm GeV}$. The requirement of successful ``flavoured''
leptogenesis in the latter case leads to the following lower limits on
$|\sin\theta_{13}\sin\delta|$, and thus on $\sin\theta_{13}$ and on
the rephasing invariant $J_{\rm CP}$ which controls the magnitude of
CP-violation effects in neutrino oscillations: $|\sin\theta_{13}\sin
\delta|,\sin\theta_{13}\gtap (0.04 - 0.09)$, $|J_{\rm CP}| \gtap
(0.009 - 0.020)$, where the precise value of the limit within the
intervals given depends on the ${\rm sign}(R_{11}R_{13})$ (or ${\rm sign}(R_{12}R_{13})$) 
and on $\sin^2\theta_{23}$.

\begin{figure}[t!!]
\begin{center}
\vspace{-1.0cm}
\begin{tabular}{cc}
\includegraphics[width=8truecm,height=6.5cm]{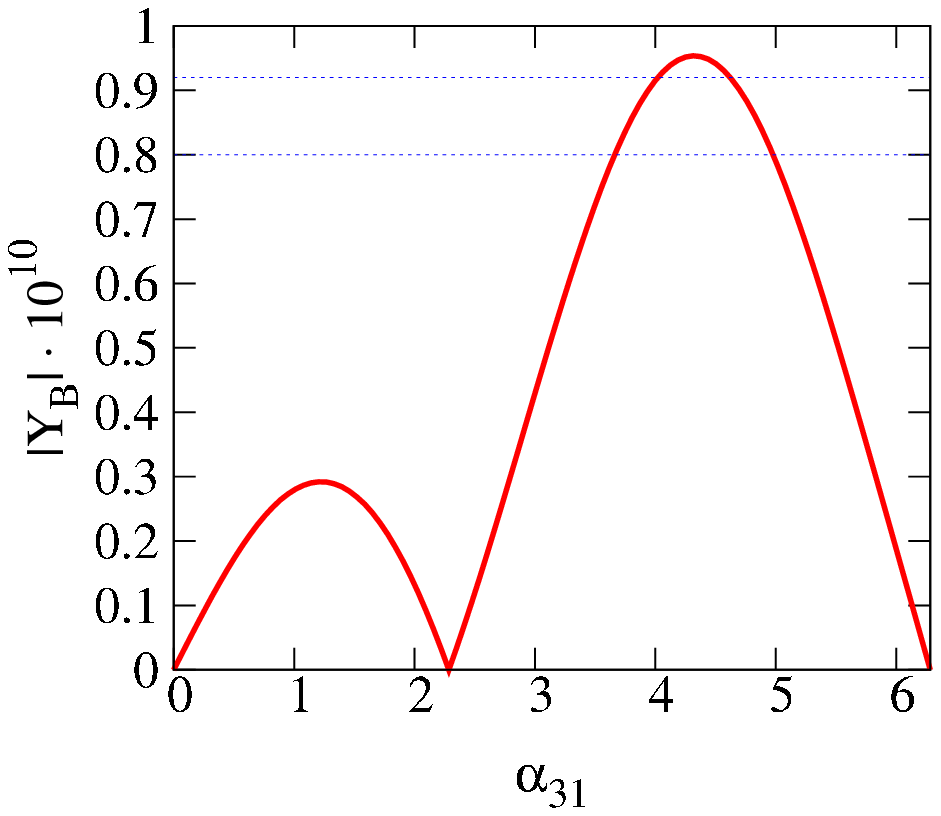}&
\includegraphics[width=8truecm,height=6.5cm]{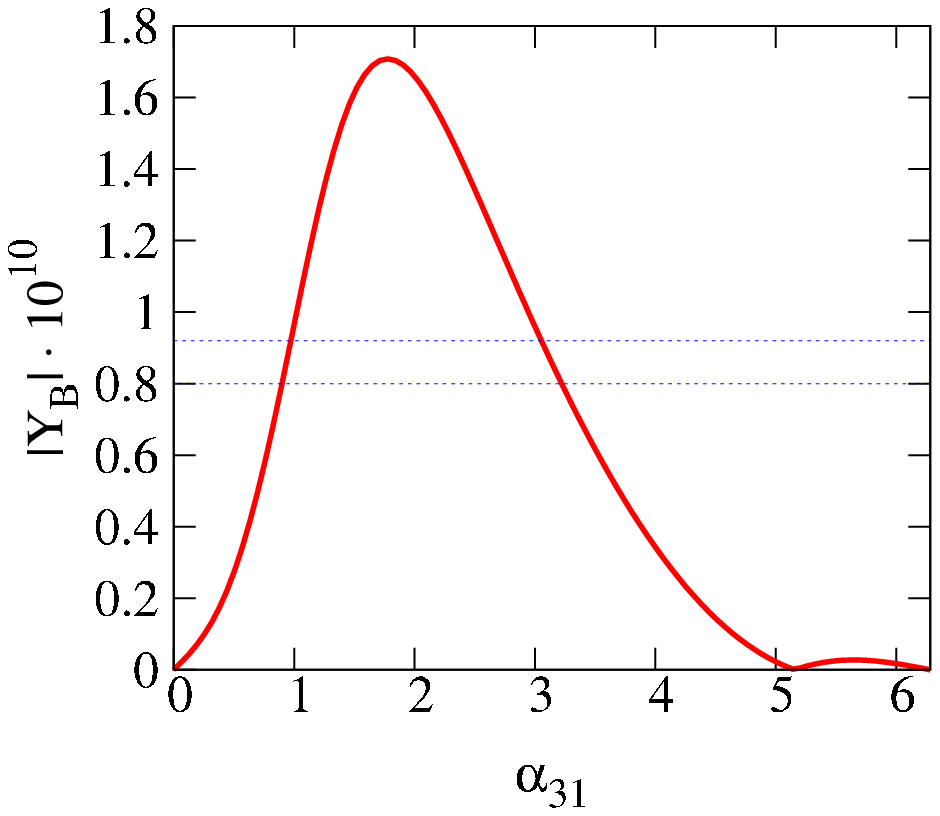}
\end{tabular}
\caption{ \label{R11_0_alpha} The dependence of $|Y_B|$ on
  $\alpha_{31}$ (Majorana CP-violation), in the case of IH spectrum,
  real $R_{1j}R_{1k}$, $R_{12}= 0$, $s_{13}=0$, $M_1=10^{11}~{\rm GeV}$, and for i)
  $|R_{11}|=0.18$, $m_3=5.6\times 10^{-3}{\rm~eV}$, ${\rm sign}(R_{11}R_{13})=+1$ (left
  panel), and ii) $|R_{11}|=0.48$, $m_3=9.3\times
  10^{-3}{\rm~eV}$, ${\rm sign}(R_{11}R_{13})=-1$ (right panel). The values of $m_3$ and $|R_{11}|$
  used maximise $|Y_B|$ at $\alpha_{31} = \pi/2$.  The horizontal dotted lines indicate the
  allowed range of $|Y_B| = (8.0 - 9.2)\times 10^{-11}$.}
\end{center}
\end{figure}

\begin{figure}[t!!]
\begin{center}
\begin{tabular}{cc}
\includegraphics[width=8truecm,height=6.5cm]{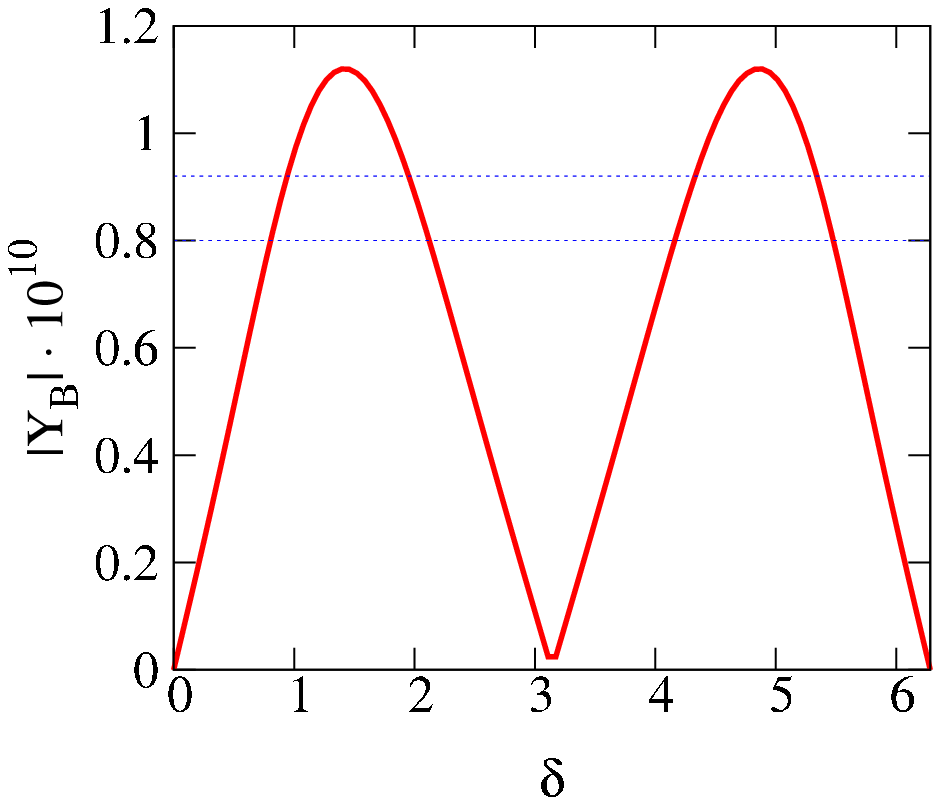}&
\includegraphics[width=8truecm,height=6.5cm]{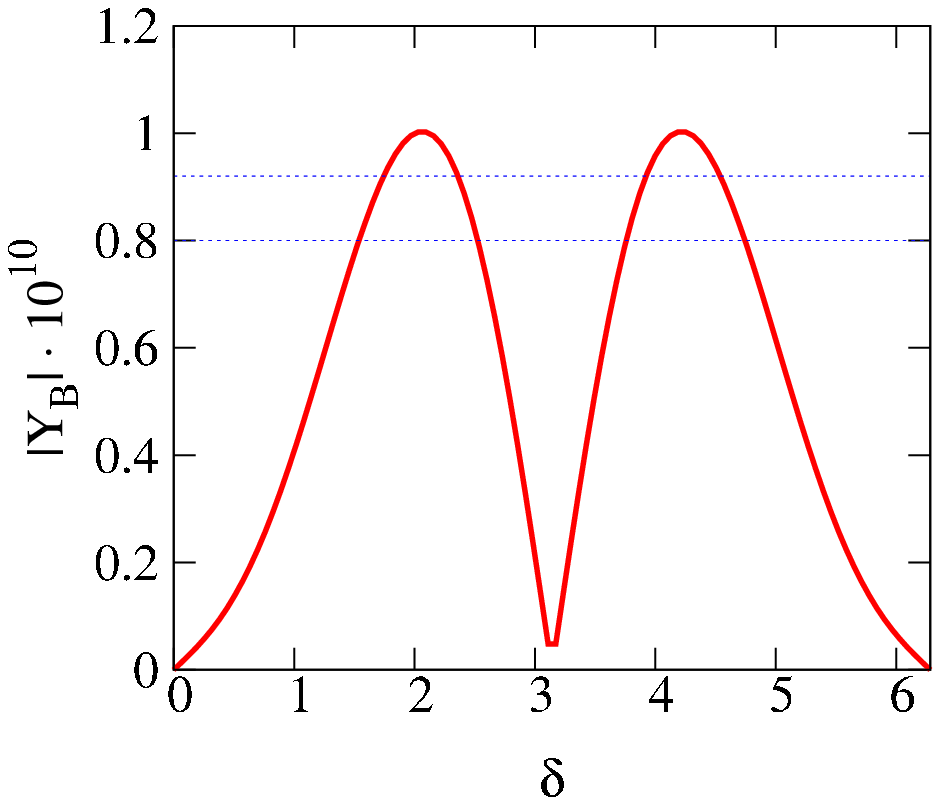}
\end{tabular}
\caption{\label{R11_R12_delta} The dependence of $|Y_B|$ on
  $\delta$ (Dirac CP-violation), in the case of IH spectrum,
  real $R_{1j}R_{1k}$, $s_{13}=0.2$ and for i) $M_1=3\times10^{11}~{\rm GeV}$, $\alpha_{32}\equiv\alpha_{31}-\alpha_{21}=0$, $R_{11}= 0$, 
  $|R_{12}|=0.29$, $m_3=6.7\times 10^{-3}{\rm~eV}$, ${\rm sign}(R_{12}R_{13})=-1$ (left
  panel), and ii) $M_1=5\times10^{11}~{\rm GeV}$, $\alpha_{31}$=0, $R_{12}= 0$, $|R_{11}|=0.22$, $m_3=8.6\times
  10^{-3}{\rm~eV}$, ${\rm sign}(R_{11}R_{13})=+1$ (right panel).  The values of $m_3$ and $|R_{1j}|$, $j=1,2$,
  used maximise $|Y_B|$ at $\delta = \pi/2$.  The horizontal dotted lines indicate the
  allowed range of $|Y_B| = (8.0 - 9.2)\times 10^{-11}$. }
\end{center}
\end{figure}

In Fig \ref{R11_0_alpha} the dependence of the generated baryon asymmetry with respect to the 
Majorana phase $\alpha_{31}$ is reported. The case analysed corresponds to the limit $R_{12}\cong 0$, 
for which condition (\ref{limit}) is fulfilled and the lightest neutrino mass $m_3$ gives an important contribution
to the corresponding CP-asymmetry in the range $5\times 10^{-6}~{\rm eV} \ltap m_3 \ltap
5\times 10^{-2}$ eV. The neutrino mixing angle $\theta_{13}$ is set to zero and no CP-violation
arising from the Dirac phase is assumed. The values of $m_3$ and $|R_{11}|$ are computed in order to maximise the
baryon asymmetry $|Y_B|$ at $\alpha_{31}=\pi/2$ for the two possible choices: ${\rm sign}(R_{11}R_{13})=\pm 1$.

The case in which the observed baryon asymmetry arises from the CP-violating contribution
of the Dirac phase $\delta$ is presented in the two plots in Fig \ref{R11_R12_delta}. We set
$\sin\theta_{13}=0.2$ and $R_{11}\cong 0$, $R_{12}\cong 0$ in the left and right panel, respectively.
The effective Majorana phase is CP-conserving in both cases. The values of the lightest neutrino mass
and the elements $|R_{1j}|$, $j=1,2$, are chosen in such a way as to maximise the baryon asymmetry
$|Y_B|$. We remark that a value of $\delta\neq 0, \pi$, Majorana CP-conserving phases $\alpha_{31}=\pi$ and $\alpha_{21}=0$ and real matrix $R$ imply ``low energy'' CP-violation due to the
neutrino mixing matrix, but a ``high energy'' contribution associated with the matrix $R$ is still present,
as already discussed in section 2. For this reason we didn't consider this possibility in the reported results.

\section{Light Neutrino Mass Spectrum with Normal Ordering and Real $R_{1j}$}

A light neutrino mass spectrum with normal 
ordering (hierarchy) gives very different predictions with respect 
to the previous one. The case of negligible $m_1$ and real
(CP-conserving) elements $R_{1j}$ of $R$ was analysed in detail
in~\cite{PPRio106}. In searching for possible significant effects of non-negligible $m_1$
in leptogenesis we have considered values of $m_1$ as large as 0.05
eV, $m_1 \leq 0.05$ eV.  For $3\times 10^{-3}~{\rm eV} \ltap m_1 \ltap
0.10$ eV, the neutrino mass spectrum is not hierarchical; the spectrum
exhibits partial hierarchy (see, e.g. \cite{BPP1}), i.e.  we have $m_1
< m_2 < m_3$. 

\begin{figure}[t!!]
\begin{center}
\vspace{-1cm}
\includegraphics[width=13.5cm,height=9.5cm]{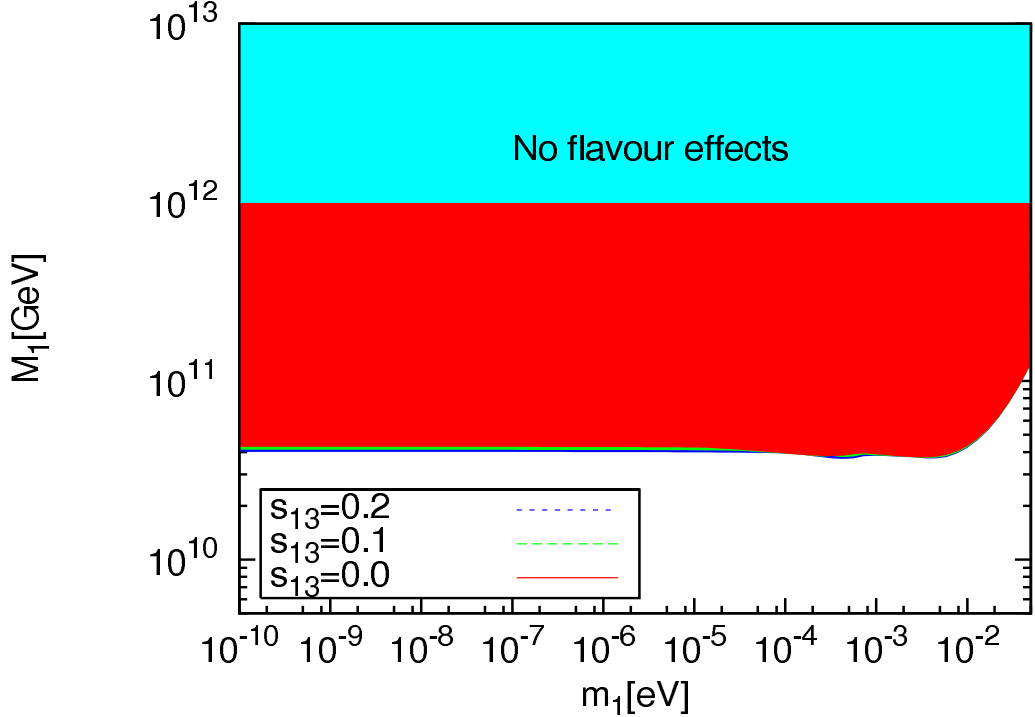}
\caption{
  \label{fig1NOM1m1} Values of $m_3$ and $M_1$ for which the
  ``flavoured'' leptogenesis is successful, generating baryon
  asymmetry $|Y_B| = 8.6\times 10^{-11}$ (red/dark shaded area). The
  figure corresponds to hierarchical heavy Majorana neutrinos, light
  neutrino mass spectrum with inverted ordering (hierarchy), $m_3 <
  m_1 < m_2$, and real elements $R_{1j}$ of the matrix $R$. The results shown are obtained
  using the best fit values of neutrino oscillation parameters:
  $\deltasol = 8.0\times 10^{-5}~{\rm eV}^2$, $\deltaatm = 2.5\times
  10^{-3}~{\rm eV}^2$, $\sin^2\theta_{12}=0.30$ and
  $\sin^22\theta_{23}=1$.}
\end{center}
\end{figure}

These results are illustrated in Fig.~\ref{fig1NOM1m1}, showing the
correlated values of $M_1$ and $m_1$ for which one can have successful
leptogenesis. The figure was obtained using the same general method of
analysis we have employed to produce Fig.~\ref{fig1IOM1m3}.  For given
$m_1$ from the interval $10^{-10} \leq m_1 \leq 0.05$ eV, a thorough
scan of the relevant parameter space was performed in the calculation
of $|Y_B|$, searching for possible non-standard features (enhancement
or suppression) of the baryon asymmetry.  The real elements $R_{1j}$
of interest of the matrix $R$, were allowed to vary in their full
ranges determined by the condition of orthogonality of $R$: $R^2_{11}
+ R^2_{12} + R^2_{13} = 1$. The Majorana and Dirac phases
$\alpha_{21,31}$ and $\delta$ were varied in the interval $[0,2\pi]$.
The calculations were performed again for three values of the CHOOZ
angle, $\sin\theta_{13} = 0;~0.1;~0.2$.  The relevant heavy Majorana
neutrino mass $M_1$ was varied in the interval $10^{9}~{\rm GeV} \ltap
M_1 \ltap 10^{12}$ GeV.  For given $m_1$, the minimal value of the
mass $M_1$, for which the leptogenesis is successful generating $Y_B
\cong 8.6\times 10^{-11}$, was obtained for the values of the other
parameters which maximise $|Y_B|$. The ${\rm min}(M_1)$ thus
calculated did not show any significant dependence on $s_{13}$.  For
$m_1 \ltap 7.5\times 10^{-3}$ eV we did not find any noticeable effect
of $m_1$ in leptogenesis: the results we have obtained practically
coincide with those corresponding to $m_1 = 0$ and derived in
\cite{PPRio106}.  For $7.5\times 10^{-3}~{\rm eV} \ltap m_1 \leq 5\times 10^{-2}$ eV 
the predicted baryon asymmetry $Y_B$ for
given $M_1$ is generically smaller with respect to the asymmetry $Y_B$
one finds for $m_1 = 0$.  Thus, successful leptogenesis is possible
for larger values of ${\rm min}(M_1)$.  The corresponding suppression
factor increases with $m_1$ and for $m_1 \cong 5\times 10^{-2}$ eV
values of $M_1 \gtap 10^{11}$ GeV are required.

The results we have obtained for light neutrino mass spectrum with
normal ordering can vary significantly if one of
the elements $R_{1j}$ is equal to zero. In particular, if $R_{11} \cong 0$, we did not find any
significant enhancement of the baryon asymmetry $|Y_B|$, generated
within ``flavoured'' leptogenesis scenario with real matrix $R$ and
CP-violation provided by the neutrino mixing matrix $U_{\rm PMNS}$,
when the lightest neutrino mass was varied in the interval
$10^{-10}~{\rm eV}\leq m_1 \leq 0.05~{\rm eV}$. If, however, $R_{12} \cong 0$, the
dependence of $|Y_B|$ on $m_1$ exhibits qualitatively the same
features as the dependence of $|Y_B|$ on $m_3$ in the case of neutrino
mass spectrum with inverted ordering (hierarchy), although ${\rm max}(|Y_B|)$ is somewhat
smaller than in the corresponding IH spectrum cases (see \cite{MPST} for a detailed discussion on this point).
As a consequence, it is possible to reproduced the observed value of $Y_B$
if the CP-violation is due to the Majorana phase(s) in $U_{\rm PMNS}$
provided $M_1\gtap 5.3\times 10^{10}~{\rm GeV}$.

\section{Conclusions}

The analysis we have performed in \cite{MPST} shows that within the thermal ``flavoured'' leptogenesis scenario, 
the value of the lightest neutrino mass can have non negligible effects on
the magnitude of the baryon asymmetry of the Universe in the cases of light neutrino mass spectrum with
inverted and normal ordering (hierarchy). In particular, as regards the IH spectrum, 
one can have an enhancement of the baryon asymmetry by a factor of $\sim 100$  with respect to the value 
corresponding to $m_3\cong 0$, thus allowing for the generation of a matter-antimatter asymmetry
compatible with the experimental observation.

\section*{Acknowledgements}
 
This work was supported in part by the INFN and Italian MIUR programs on
 ``Fisica Astroparticellare''.

\end{document}